%
%

\input amstex
\documentstyle{conm-p}

\issueinfo{00}
{}
{}
{2007}


\topmatter
\title From Stationary Phase to Steepest Descent\endtitle
\author Spyridon Kamvissis\endauthor

\noindent

\address Department of Applied Mathematics, University of Crete,  Knossos, Greece \endaddress
\email spyros\@tem.uoc.gr\endemail

\address \endaddress


\thanks \endthanks

\date \enddate

\dedicatory Dedicated to Percy Deift on his 60th birthday\enddedicatory


\endtopmatter

\document

\define\[{\left[}%
\define\]{\right]}%
\define\({\left(}%
\define\){\right)}%
\baselineskip=20pt

\bigskip

0. INTRODUCTION

\bigskip

The so-called nonlinear stationary-phase-steepest-descent method for the 
asymptotic analysis
of Riemann-Hilbert factorization problems has been very successful in providing

(i) rigorous results on long time, long range and semiclassical asymptotics 
for solutions of completely integrable equations and correlation functions of exactly
solvable models, 

(ii) asymptotics for orthogonal polynomials of large degree,

(iii) the eigenvalue distribution of random
matrices of large dimension (and related  universality results), 

(iv) proofs of important results in combinatorial probability
(e.g. the limiting distribution of the length of longest increasing subsequence
of a permutation, under uniform distribution).

Even though the stationary phase idea was first applied to a Riemann-Hilbert
problem and
a nonlinear integrable equation by Its ([I], 1982)
the method became systematic and rigorous in the work of Deift and Zhou ([DZ], 1993).
As a recognition of the fruitfulness of the method,
Percy Deift was invited to give a plenary address to the recent ICM
in Madrid, on the subject of universality in mathematics and physics. 
Of course,
the main mathematical tool in proving universality theorems has  been the
nonlinear stationary-phase-steepest-descent method.

In analogy to the linear stationary-phase and steepest-descent methods,
where one asymptotically reduces the given exponential integral  to
an exactly solvable one, in the nonlinear case
one asymptotically reduces the given Riemann-Hilbert problem to
an exactly solvable one.

Our aim here is to clarify  the distinction between the 
stationary-phase idea and the steepest-descent idea, stressing
the importance of actual steepest-descent contours in some problems.
We claim that the distinction partly mirrors the
self-adjoint / non-self-adjoint dichotomy  of the underlying Lax operator.
To this aim we first have to review some of the main groundbreaking ideas 
(due to 
Percy Deift and his collaborators)
appearing in the self-adjoint case;  then we describe  recent results ([KMM], [KR])
in the non-self-adjoint case, that we  see  as a natural extension.
We mostly use the defocusing / focusing nonlinear Schr\"odinger equation as
our working model, but we also digress to the KdV at some point.

We stress both here and in the main text that
an extra  feature appearing only in the nonlinear asymptotic theory
is the Lax-Levermore variational problem, discovered in 1979, before the work
of Its, Deift and Zhou, but closely related to the so-called
"g-function" which is catalytic in the  process of deforming Riemann-Hilbert
factorization problems to exactly solvable ones.

\bigskip

1. THE LINEAR   METHOD

\bigskip

Suppose one considers the Cauchy problem for the
linearized KdV equation:  $u_t + u_{xxx}=0$.  It can of course
be solved via Fourier transforms. The end result of the Fourier method
is an exponential integral.
To understand the long time asymptotic behavior of the integral  one needs to 
apply the stationary-phase  method (see e.g.[E]). The underlying principle,
going back to Stokes and Kelvin, is that the
dominating contribution comes from the vicinity of 
the $stationary~phase~points$.
Through a local change of variables at each
stationary phase point and
using integration by parts we can calculate each
contributing integral asymptotically to all orders with exponential error.
It is essential here that the phase \footnote{$\xi$
is the spectral variable} $x \xi- \xi^3 t$
is real  and that the
stationary phase points are real.

On the other hand, suppose we have something like the Airy
exponential integral
$$\aligned
Ai(z) = {1 \over \pi} \int_0^{\infty} cos ({s^3 \over 3} + zs) ds,
\endaligned
\tag1
$$
and we are interested in $z \to \infty$. Set $s = z^{1/2} t$ and
$x = z^{3/2} $.
$$
\aligned
Ai(x^{2/3}) = {x^{1/3} \over {2\pi}} \int_{-\infty}^{\infty} exp (ix ({t^3 \over 3} + t)) dt.
\endaligned
\tag2
$$
The phase is $h(t) =  {t^3 \over 3} + t$ 
and the zeros of
$h'(t) =  (t^2 + 1)$ are $\pm i$.
As they are not real, and since the integrand is analytic, one must
deform the integral off the real line, along particular paths.  These
are referred to as {\it steepest descent paths}.
They are given by the simple characterization
$$\aligned
Im h(t) = constant.
\endaligned
\tag3
$$
In our particular example,
the curves of steepest descent are the imaginary axis 
and the two branches of a hyperbola.
By deforming to one of these branches, 
we finally end up with 
with integrals which can be analyzed
directly, using the so-called Laplace's method (which is simpler than
the stationary phase method). We thus recover asymptotics valid to all orders.

The nonlinear method generalizes the ideas above, but also employs  new ones.

\bigskip

2. THE  NONLINEAR METHOD

\bigskip

(i) The Stationary Phase Idea

\bigskip

Consider the 
defocusing nonlinear Schr\"odinger equation
$$
\aligned
i\partial_t\psi +
\partial_x^2\psi - |\psi|^2\psi = 0, \\
under~~~~\psi(x,0)=\psi_0(x),
\endaligned
\tag4
$$
where the initial data function lies in, say, Schwartz space.
The analog of the Fourier transform is  the scattering coefficient
$r(\xi)$ for the Dirac operator
$$
\aligned
L= \pmatrix &i  \partial_x & i \psi_0(x) \\ &-i \psi_0^* (x) &-i \partial_x
\endpmatrix.
\endaligned
$$
Suppose we are now interested in the long time  behavior of
the solution to (4).
The inverse scattering problem 
can be posed in terms of a Riemann-Hilbert factorization problem.

THEOREM. There exists a
2x2 matrix $Q$ with analytic entries in the upper and lower open half-planes,
such that the normal limits $Q_+, Q_-$, as $\xi$ approaches the real line from
above or below respectively, exist and satisfy
$$
\aligned
Q_+(\xi) = Q_-(\xi) \pmatrix
&1 - |r(\xi)|^2
&- r^*(\xi) e^{-2i\xi x -4i \xi^2 t}
\\  & r(\xi) e^{2i\xi x + 4i \xi^2 t} &1
\endpmatrix, ~~~~~~~~~~Im \xi =0,\\
and~~~~~ lim_{\xi \to \infty} Q (\xi) =I.
\endaligned
\tag5
$$
The solution to (4) is recovered via
$$
\aligned
\psi(x,t) = -2 lim_{\xi \to \infty} \xi Q_{12} (\xi).
\endaligned
\tag6
$$
It was first realized by Its [I, IN],
that the leading order behavior of the long time asymptotics for the solution of (4)
can be described by replacing the problem (5) by a 
"local" model Riemann-Hilbert problem
located in a small neighborhood of the $stationary~phase~point~\xi_0=-{ x \over {4t}}$ 
satisfying 
$\Theta' (\xi_0) =0$ where $\Theta = \xi x + 2 \xi^2 t$, or, equivalently to a problem similar to
(5) but where $\xi$ is replaced by the constant value $\xi_0$
(see (7) below).  But no idea was given on how to show
that this solution of the full problem and of the model problem are actually
close to each other.  To show how to do this, was the work of Deift and Zhou.
\footnote{See [DZ] for a comprehensive review of the history of the problem before the 
use of Riemann-Hilbert problem tecniques.}
The  basic ideas of [DZ] have been used in
all works on the stationary-phase-steepest-descent-method since.
They include:

1. Appropriate lower/diagonal/upper factorizations of jump matrices.

2. Equivalence of the solvability of  inhomogeneous
matrix Riemann-Hilbert problems to
the invertibility of  associated singular integral operators.
This idea goes at least back to Gohberg [CG]. A crucial
contribution of Beals and Coifman [BC] was to make this
precise for contours with sef-intersections (which they needed 
in their study of the
inverse scattering of first order systems).
These ideas where further developed by Zhou [Z] who provided 
a very useful  existence theorem for matrix 
Riemann-Hilbert problems with jumps and jump contours satisfying some
special Schwarz reflection type symmetries and an 
integral formula expressing the solution
of the  Riemann-Hilbert problem in terms of the inverse of a
particular   weighted Cauchy  operator depending on  a given
factorization of the jump matrix and thus taking advantage
of the factorization mentioned above. Perturbing Zhou's formula
provides a nice way to show that
under some conditions, small changes in the jump data result in small changes in the solution.

3. Introduction and solution of auxiliary scalar problems.

\bigskip
Following analyticity and the above ideas one ends up with a problem on a $small~
cross~centered~at~the~stationary~phase~point$.
Using a rescaling
the Riemann-Hilbert problem is   rescaled
to a new problem on an infinite cross.
After deforming the components of the cross back to the real line,
it is equivalent to the following problem on the real line:
$$
\aligned
H_+(\xi) = H_-(\xi)exp (-i\xi^2 \sigma_3) \pmatrix
&1 - |r(\xi_0)|^2
&- r^*(\xi_0)
\\  & r(\xi_0)  &1 \endpmatrix
exp (i\xi^2 \sigma_3),\\
H(\xi) \sim \xi^{i \nu \sigma_3},
\endaligned
\tag7
$$
where $\nu$ is a constant depending only on $\xi_0$ and $\sigma_3= \pmatrix
&1&0 \\& 0 &-1 \endpmatrix$ is a Pauli matrix. So the jump matrix
$\pmatrix
&1 - |r(\xi)|^2
&- r^*(\xi)
\\  & r(\xi)  &1 \endpmatrix$ of the original problem is replaced by its value at $\xi_0$.

Problem (7) can be solved explicitly. 
Written in terms of the new unknown
$ H(\xi)exp (-i\xi^2 \sigma_3)$ it has a constant jump
and can thus be reduced to a first order linear matrix ODE ([I]).

\bigskip

(ii). The finite-gap g-function mechanism and a "shock" phenomenon with no linear analogue.

\bigskip

An important step halfway between the leap from the "stationary phase" idea
to the general definition of a "steepest descent contour" is the introduction of
the so-called finite-gap  g-function mechanism. The g-function was  introduced in 
[DZ95] in the special case of genus 0 and  in [DVZ94] in the
special case of genus 1 but
the full force of the finite-gap g-function idea and the connection to the Lax-Levermore
variational problem was first  explored in the analysis of
the KdV equation [DVZ97]
$$
\aligned
u_t - 6u u_x + \epsilon^2 u_{xxx} =0,\\
u(x,0)=u_0(x),
\endaligned
$$
in the limit as $\epsilon \to 0$.
Assume for simplicity, that the initial data are real analytic, positive and consist of
a "hump" of unit height.

The associated RH problem is
$$
\aligned
S_+(z) = S_-(z) \pmatrix
&1 - |r(z)|^2
&- r^*(z) e^{{-iz x -4i z^3 t}\over \epsilon}
\\  & r(z) e^{{iz x + 4i z^3 t} \over \epsilon} &1
\endpmatrix, ~~~~~~~~~~Im z =0,\\
and~~~~~ lim_{z \to \infty} S (z) =(1,1),
\endaligned
$$
where $r$ is the reflection coefficient for the Schr\"odinger
operator with potential $u_0$.
The solution of KdV is recovered via
$$
\aligned
u(x,t; \epsilon) = -2 i\epsilon {{\partial} \over {\partial x}}   S_1^1 (x,t; \epsilon),  \endaligned
$$
where $ S_1^1$ is the resdiue of the first entry of $S$ at infinity.
The reflection coefficient $r$ also depends on $ \epsilon$.
In fact, the WKB approximation is
$$
\aligned
r(z) \sim -i e^{{-2i \rho(z) }\over \epsilon} \chi_{[0,1]} (z)\\
1- |r(z)|^2 \sim e^{{-2  \tau(z) }\over \epsilon},
\endaligned
$$
where
$$
\aligned
\rho(z) = x_+ z +  \int_{x_+}^{\infty} [z-(z^2 -u_0(x))^{1/2}] dx,\\
\tau(z) =  Re \int (u_0(x)- z^2)^{1/2} dx 
\endaligned
$$
and $x_+(z)$ is the largest  solution of $u_0(x_+) = z^2$.

[DVZ97] introduce the following change of variables
$ \hat S(z) = S(z) e^{i{g(z) \sigma_3  } \over \epsilon} $ 
where $g$ is a scalar function defined  by the following conditions.

1. $g$ is analytic off the interval $[0, 1] $, the normal limits $g_+, g_-$ of $g$ exist
along $[0,1]$ and $g$ vanishes at infinity.

2. "Finite gap ansatz". Define $h(z) = g_+(z) + g_-(z) -2 \rho + 4tz^3 +xz$.
There exists a finite set of disjoint open real intervals ("bands")
$I_j \in [0, 1] $ such that 

3a. For $ z \in \cup_j I_j,$ we have 
$-\tau < (g_+-g_-) /2i <0$ and $h'=0$.

3b. For $ z \in [0, 1] \setminus \cup_j I_j,$ we have
$2i\tau =g_+-g_-  $ and $h' <0$.

The conditions above are meant to 
determine not only $g$ but also the band-gap structure in $[0,1]$.
In general (for any data $u_0$) it is not true that the above conditions can be satisfied.
It is believed  however  that under the condition of analyticity a g-function
satisfying the "finite gap ansatz" exists. (In fact  [K00] gives  a proof of the 
"finite gap ansatz" in the analogous problem of the continuum Toda equations.) 
Assuming that there is a g-function satisfying the three conditions above
one can show that the RH problem reduces to one
supported on the bands $I_j$ with jumps of the form
$$
\aligned
\pmatrix
&0
&-ie^{-ih(z)/\epsilon}\\
&-ie^{ih(z)/\epsilon} &0
\endpmatrix,
\endaligned
$$
and in fact, because of (iib), $h(z)$ is a real constant on each band $I_j$.
This RH problem can be solved explicitly via theta functions. The details in 
[DVZ97] involve the so-called "lens"-argument: auxiliary contours are introduced near
pieces of the real line (one below and one above each band/gap)
and appropriate factorizations and analytic extensions are used,
very similarly to the subsection above. The conditions for $g$ above are chosen precisely to
make the lens argument work.

As is remarked in [DVZ94] the fact that the new RH problem is on  slits
"is a new and essentially nonlinear feature of our nonlinear stationary phase method".
Unlike [DVZ94] where it is defined explicitly via an integral formula, in [DVZ97]
the g-function is only defined implicitly via the conditions above, which may or may not
admit a solution.

\bigskip

(iii) The Lax-Levermore Variational Problem

\bigskip

The g-function satisfying conditions (i), (ii), (iia), (iib) can be written as
$$
\aligned
g(z) = \int log (z-\eta) d\mu (\eta)
\endaligned
$$
where $\mu$ is a continuous measure supported in $\cup_j I_j$.
In a sense, the reduction of the given RH problem to an explicilty solvable one
depends on the existence of a particular measure. Conditions (i), (ii), (iia), (iib)
turn out to be equivalent to a maximization problem for logarithmic potentials
under a particular external field depending on $x,t, u_0(x)$  over positive  measures
with an upper constraint. This is related to the famous 
Lax-Levermore Variational Problem [LL]. Even though it appears as an afterthought
in [DVZ97] (though clearly serving as inspiration), 
it seems that its analysis is essential for the justification 
of the method (as in [K00]).

\bigskip

3. STEEPEST DESCENT CONTOURS 

\bigskip

Having reviewed some essential  ideas in the previous sections, we are ready to
consider  the focusing NLS equation, following [KMM].
$$
\aligned
i\hbar\partial_t\psi +
\frac{\hbar^2}{2}\partial_x^2\psi + |\psi|^2\psi = 0, \\
under~~~~data~~\psi(x,0)=\psi_0(x).
\endaligned
$$
Note that the Lax operator
$$
\aligned
L= \pmatrix &i h \partial_x &-i \psi_0(x) \\ &-i \psi_0^* (x) &-ih \partial_x
\endpmatrix,
\endaligned
$$
is non-self-adjoint.
We shall see that the deformation of the semiclassical RH problem can be no more
confined to a small neighborhood of the real axis  but is instead fully two-dimensional.
A $steepest~descent~contour$ needs to be discovered!
\footnote{
By the way, in the
long time asymptotics for the above  with 
$\hbar =1$  a collisionless shock phenomenon is also present; for $x,t$ in the
shock region the deformed  RH problem is 
supported on a vertical imaginary slit. (See [K96].)
But here,
we rather focus on the semiclassical problem $\hbar \to 0$ which is far
more complicated.}

For simplicity consider the very specific data 
$\psi_0(x) = Asechx$ where $A >0$.
Let $x_- (\eta) <x_+(\eta) $ be the two solutions of $sech^2(x) + \eta^2 =0$.
Also assume that $\hbar = A/N$ and consider the limit $N \to \infty$.
It is known that the reflection coefficient is identically zero and that the 
eigenvalues of $L$
lie uniformly placed on the imaginary segment $[-iA, iA]$.
In fact the eigenvalues are the 
points $\lambda_j = i\hbar (j+1/2), j=0, ..., N-1$ and their conjugates. 
The norming constants oscillate between $-1$ and $1$.

The associated RH problem is a meromorphic problem 
with no jump: to find a rational function with prescribed
residues at the poles $\lambda_j $ and their conjugates. It can be turned into a
holomorphic problem by constructing two loops, one denoted by $C$
say, encircling the $\lambda_j$ and one, $C^*$, encircling their conjugates. 
We redefine the unknown 2x2 matrix inside the loops so that
the poles vanish 
(there is actually a  discrete infinity of choices, 
corresponding to an infinity of analytic interpolants of the norming constants,
see below)
and thus arrive at a nontrivial jump across the two loops,
encircling the segments $[0,iA]$ and $[-iA,0]$ respectively.
This is a trivial deformation, valid for any $h$ (not necessarily small).
The discrete nature of the spectrum of $L$ is mirrored in the discrete
nature of the jump matrices: they involve a logarithmic integral with
respect to a discrete measure. We sometimes refer to this as a discrete
Riemann-Hilbert problem.

THEOREM. Let
$ d\mu = (\rho^0 (\eta) + (\rho^0)^* (\eta^*)) d\eta$,
where $\rho^0=i $ is the asymptotic density of eigenvalues
supported on the linear segment $[0,iA]$. Set
$ X (\lambda) =  \pi ( \lambda -iA) .$

Letting $M_+ $ and $M_-$ denote the limits of
$M$  on $\Sigma = C \cup C^*$ from  left and right
respectively, we define the Riemann-Hilbert factorization problem
$$
\aligned
M_+(\lambda) = M_-(\lambda) J(\lambda) ,\\
where~~~\\J(\lambda) =
v(\lambda), \lambda \in C,\\
= \sigma_2 v(\lambda^*)^* \sigma_2, \lambda \in C^*,\\
lim_{\lambda \to \infty} M(\lambda) =I,
\endaligned
$$
and
$$
\aligned
v(\lambda)=
\pmatrix
&1 &  -i~exp ({1 \over h} \int log(\lambda -\eta) d\mu(\eta))
exp (-{1 \over h} (2i\lambda x + 2i \lambda^2 t - X(\lambda)))\\
&0 &1
\endpmatrix.
\endaligned
\tag8
$$

Then the solution of the initial value problem for the focusing NLS equation  is given by
$\psi(x,t) = 2i~lim_{\lambda \to \infty} (\lambda M_{12})$.

Note that in the statement of the theorem the measure in the logarithmic integral 
is now continuous. We have effectively substituted a discrete set of eigenvalues
by its continuous limiting density. This is only valid as $h \to 0$ and the
rigorous justification of the 
discrete-to-continuum passage is far from trivial, especially near the
point $0$ where the loops $C, C^*$ hit the eigenvalue spike.

The analysis in [KMM] makes use of all 
the ideas described in the previous sections
(factorization, lenses, the weighted Cauchy operator, 
an auxiliary scalar problem), but it
also takes care of the fact that while
the loops can be deformed anywhere away from the poles as long as $h$ is not small,
they have to be eventually located at a
very specific position in order to asymptotically simplify the
RH problem, as $h \to 0$. Appropriately, the
definition of a g-function has to be generalized. Not only
will it introduce  the division of the  loop into
arcs, called "bands" and "gaps", but it must implicitly select a contour.
Rather than giving the complicated set of equations and inequalities defining the
g-function, we will rather focus on the associated
variational problem; it is not a maximization problem but
rather a maximin problem.
Here's the setting.

Let
$ \Bbb H = \{ z: Im z >0 \} $ be the complex upper-half plane  and
 $\bar \Bbb H =  \{ z: Im z \geq 0 \} \cup \{\infty\}$
 be the closure  of $ \Bbb H $. Let
also
$ \Bbb K = \{ z: Im z >0 \} \setminus \{ z: Rez =0, 0< Im z \leq A \}$.
In the closure of this space, $\bar \Bbb K $, we consider the points
$ix_+$ and $ix_-$, where $0 \leq x < A$ as distinct.
In other words, we cut a slit in the upper half-plane along the
segment $(0, iA)$ and distinguish between the two sides of the slit.
The point infinity belongs to $\bar \Bbb K$, but not $\Bbb K$.
Define
$G(z; \eta)$ to be the Green's function for the upper half-plane
$$
\aligned
G(z; \eta) = log {{ |z-\eta^*| } \over {|z-\eta|}}
\endaligned
$$
and  let $d\mu^0 (\eta)$ be the nonnegative measure $-i d\eta$
on the segment $[0,iA]$ oriented from 0 to iA. The star denotes
complex conjugation. Let the "external field" $\phi$ be defined by
$$
\aligned
\phi (z) =
-\int G(z; \eta) d\mu^0(\eta) - Re (\pi  (iA-z) +2i  (z x + z^2 t) ),
\endaligned
$$
where, without loss of generality $x >0$.

Let $\Bbb M$ be  the set of all positive Borel measures on $\bar \Bbb K$,
such  that both the free energy
$$
\aligned
E(\mu) = \int \int G(x,y) d\mu(x) d\mu(y), ~~~\mu \in \Bbb M
\endaligned
$$
and $\int \phi d\mu$ are finite.
Also, let
$$
\aligned
V^{\mu} (z) = \int G(z,x) d\mu(x), ~~~\mu \in \Bbb M.
\endaligned
$$
be the Green's potential of the measure $\mu$.
The weighted energy of the field $\phi$ is
$$
\aligned
E_{\phi} (\mu) =  E(\mu) + 2 \int \phi d\mu,
\endaligned
$$
for any $\mu \in \Bbb M$.

Now, given any curve $F$ in $\bar \Bbb K$, the equilibrium measure
$\lambda^F$ supported in $F$ is defined by
$$
\aligned
E_{\phi} (\lambda^F) = min_{\mu \in  M(F)} E_{\phi} (\mu),
\endaligned
$$
where $M(F)$ is the set of measures in $\Bbb M$ which are supported in $F$, 
provided such a measure exists.

It turns out that
the finite gap ansatz is equivalent to the existence of a so-called  S-curve
joining the points $0_+$ and $0_-$ and lying entirely in $\bar \Bbb K $.
By S-curve we mean  an oriented  curve $F$ such that the equilibrium measure
$\lambda^F$ exists, its support consists of a finite
union of analytic arcs 
and
at any interior point of $supp\mu$ the so called S-property is 
satisfied\footnote{${d \over {d n_+}}, {d \over {d n_-}}$ are the normal outward
derivatives on each side respectively}
 $$
\aligned
{d \over {d n_+}} (\phi + V^{\lambda^F}) =
{d \over {d n_-}}  (\phi + V^{\lambda^F}),
\endaligned
$$
The appropriate variational problem is: seek 
a "continuum"\footnote{ a compact connected set containing
$0_+, 0-$}  $C$ such that
$$
\aligned
E_{\phi} (\lambda^C)= 
max_{F \in \Bbb F}~E_{\phi} (\lambda^F) = ~max_{F \in \Bbb F}~~min_{\mu \in  M(F)} E_{\phi} (\mu),
\endaligned
\tag9
$$
where $\Bbb F$ is the set of continua lying in $\bar \Bbb K$.
The existence of a nice S-curve follows from the existence of a continuum $C$ maximizing the 
equilibrium measure, in particular the associated Euler-Lagrange equations and inequalities.

Problem (9) is the non-self-adjoint analogue of the Lax-Levermore problem and a
nonlinear analogue of (3).

\bigskip

4. JUSTIFICATION: EXISTENCE OF THE STEEPEST DESCENT PATH 

\bigskip

EXISTENCE THEOREM [KR]. For the external field $\phi$,
there exists a continuum  $F \in \Bbb F$ such that the equilibrium measure
$\lambda^F$ exists and
$$
\aligned
E_{\phi} [F] (= E_{\phi} (\lambda^F)) =
max_{F \in \Bbb F} min_{\mu \in M(F)} E_{\phi} (\mu).
\endaligned
$$

REGULARITY THEOREM [KR].
The continuum $F$  is in fact an S-curve, so long as it does not
touch the spike $[0,iA]$ at more than a finite number of points.

If $F$ touches the spike $[0,iA]$ at more than a finite number of points,
a conceptual revision is required.  We briefly
discuss this issue  in the next section.

Here are the main ideas of the proofs. Let 
$\rho_0$ be the distance between compact sets
$E, F$ in $\bar \Bbb K$ defined as
$$
\aligned
\rho_0(E,F) = max_{z \in E} min_{\zeta \in F} \rho_0 (z,\zeta). 
\endaligned
$$
Introduce the
Hausdorff metric on the set $ I ( \bar \Bbb K   )$
of closed non-empty subsets of $ \bar \Bbb K$:
$ \rho_{\Bbb K} (A,B) = sup  ( \rho_0 (A,B), \rho_0 (B,A) ).  $

Compactness of $\Bbb F$ is the necessary 
first ingredient to prove existence of a maximizing contour.
The second ingredient is semicontinuity of the energy
functional that takes a given continuum $F$ to the equilibrium
energy
on this continuum:
$$
\aligned
\Bbb E: F \to E_{\psi} [F] = E_{\psi} (\lambda^F) = \inf_{\mu \in M(F)} (E(\mu)
+ 2 \int \psi d\mu).
\endaligned
$$

For regularity, the crucial step is

THEOREM [KR]. Let $F$ be the maximizing continuum of 
and $\lambda^F$ be the  equilibrium measure. Let $x,t$ be such that
$F$ does not touch the spike $[0, iA]$ at more than a finite number of points.
Let $\mu$ be the extension of $\lambda^F$ to the
lower complex plane via $\mu(z^*) = -\mu(z)$.
Then, if $V$ is the logarithmic potential of $\mu$,
$$
\aligned
Re (\int {{d\mu(u)} \over {u-z}} + V'(z) )^2=
Re ( V'(z))^2 - 2 Re \int {{V'(z)-V'(u)} \over {z-u}} d\mu(u)  \\
+ Re [ {1 \over z^2}  \int 2 (u+z) V'(u) ~ d\mu(u)  ] .
\endaligned
$$
PROOF: By taking variations with respect to the equilibrium measure.

It is now easy to see that the support of the equilibrium measure
of the maximizing continuum is characterized by
$$
\aligned
Re \int^z (R_{\mu})^{1/2} dz =0,
\endaligned
$$
$$
\aligned
where~~~
R_{\mu}(z)=
( V'(z))^2 - 2  \int_{supp\mu} {{V'(z)-V'(u)} \over {z-u}} d\mu(u)  \\
+ {1 \over z^2} (\int_{supp\mu} 2 (u+z) V'(u) ~ d\mu(u) ) .
\endaligned
$$
The S-property follows easily and this proves the Regularity Theorem.

It is worth mentioning here the 
recent work of Tovbis, Venakides and Zhou [TVZ],
which examines the  initial value problem for the focusing NLS
(in the semiclassical limit) under two different 
classes of initial data. Under one of these classes, no eigenvalues
exist, hence the eigenvalue spike is missing. In such a case our argument above would
prove a regularity theorem without the extra assumption on the
maximizing contour (that it does not 
touch the spike $[0, iA]$ at more than a finite number of points).

\bigskip

5. CROSSING THE EIGENVALUES BARRIER AND THE QUESTION OF SECONDARY CAUSTICS.

\bigskip

If $F$ touches the spike $[0,iA]$ at more than a finite number of points,
regularity cannot be proved as above because variations
cannot be taken. In [KR] we have included a rough idea 
on how to extend the above proof. 
A more detailed argument is forthcoming [K07].
Since a complete proof is not published yet we simply summarize our 
general plan. 

One wishes to somehow allow the contour $F$ go through the spike $[0,iA]$. One
problem arising is that (the complexification of)
the external field is not analytic across the segment $[-iA, iA]$.
What is true, however, is that $V$ is analytic in a Riemann surface consisting
of infinitely
many sheets, cut along the line segment
$[-iA, iA]$. So, the appropriate underlying space for the
(doubled up) variational problem should now be a
non-compact Riemann surface, say $\Bbb L$. Now,
compactness is the crucial element in the proof
of a maximizing continuum. But we can
indeed compactify the
Riemann surface $\Bbb L$ by mapping it to a subset of the complex plane
and compactifying the complex plane.
The other problem, of course, is whether the amended variational problem
(with the modified
field defined on the Riemann surface
and with the possibility of $F$ not enclosing all
the original eigenvalues) is still appropriate
for the semiclassical  NLS. 
The argument  goes roughly as follows:

(i) Proof of the existence of an S-curve $F$ in $\Bbb L$
along the lines of [KR].

(ii) Deformation of the original discrete Riemann-Hilbert problem to the set
$\hat F$ consisting of the  projection of  $F$
to the complex plane.
At first sight, it is clear that  $\hat F$ may not encircle the spike $[0,iA]$.
It is however possible  to append S-loops (not necessarily with respect to the same
branch of the external field) 
and end up with a sum of S-loops, such that the amended $\hat F$  $does$
encircle the spike $[0,iA]$. To see this,  suppose
there is an open interval, say $(i\alpha, i\alpha_1)$,
which lies in the exterior of $\hat F$, while 
$i\alpha, i\alpha_1 \in \hat F$. Let us assume, that $\hat F$ crosses
$[0,iA]$ along bands at $i\alpha, i\alpha_1$ (if not the situation is
similar and simpler); call these bands $S, S_1$. 
Let $\beta^-, \beta^+$ be points (considered in $\Bbb C$)
lying on $S$ to the left and right
of $i\alpha$ respectively, and at a small distance
from $i\alpha$. Similarly,
let $\beta_1^-, \beta_1^+$ be points lying on $S_{1}$ 
to the left and right
of $i\alpha_{1}$ respectively, and at a small distance
from $i\alpha_{1}$.
We will show that there exists a "gap" region including the preimages of
$\beta^-, \beta_{1}^-$ lying in the $N$th sheet for $-N$ large enough,
and similarly there exists a "gap" region including the preimages of
$\beta^+, \beta_{1}^+$ lying in the $M$th sheet for $M$ large enough,
both being  regions for which the gap inequalities hold a priori, 
irrespectively of the actual S-curve, 
depending only on the external field!

Indeed, note  that the
quantity  $Re (\tilde \phi^{\sigma} (z) )$ (which defines the variational
inequalities) is a priori bounded above  by
$-\phi(z)$. For this, see (8.8) in Chapter 8 of [KMM]; there is actually a sign error:
the right formula is
$$
\aligned
Re (\tilde \phi^{\sigma} (z) )= -\phi(z)
+ \int G(z,\eta) \rho^{\sigma} (\eta)d\eta.
\endaligned
$$
Next note that the difference of the values of the
function $Re (\tilde \phi^{\sigma} (z) )$ in consecutive sheets is
$\delta Re (\tilde \phi^{\sigma}) = \pm 2 \pi Rez$, and hence the difference of
the values at points on consecutive sheets whose image under the
projection to the complex plane is $i \eta + \epsilon$, where $\eta $ is real
and $\epsilon $ is a small (negative or positive) real,
is $\delta (Re \tilde \phi^{\sigma} )= \pm 2 \pi \epsilon$.
This means that on the left (respectively right)
side of the imaginary semiaxis, the inequality $Re ( \tilde \phi^{\sigma} (z) )<0$
will be eventually (depending
on the sheet) be valid at any given small distance to it.

Applying the theory of [KR] we join the preimages of $\beta^-$ 
and  $\beta_{1}^-$ (under the projection pf $\Bbb L $ to
$\Bbb C$) lying in the $N$th sheet and the preimages of $\beta^+$ 
and  $\beta_{1}^+$ lying in the $M$the sheet. 
Finally we connect the  preimage of $\beta^+$ 
lying in the $N$th sheet to the preimage of $\beta^-$ 
lying in the $M$th sheet,  and so on. Note that the bands are S-curves with respect
to any branch of the external field.
We thus end up with an S-loop whose projection is
covering the "lacuna" $i\alpha, i\alpha_{1}$.

The original discrete Riemann-Hilbert problem  can be  trivially deformed to a
discrete Riemann-Hilbert on the resulting  (projection of the)
union of S-loops. All this is possible even in the
case where $\hat F$ self-intersects.

(iii) Deform the discrete Riemann-Hilbert problem to the continuous one with the right
band/gap structure
(on $\hat F$; according to the projection of the  equilibrium measure on $F$), which is
then explicitly solvable via theta functions.
Both the discrete-to-continuous approximation and the opening of the lenses needed
for this deformation are justified  as in [KMM] (see also the article [LM] mentioned below 
for the delicate study of the Riemann-Hilbert problem near the points where  
$\hat F$ crosses the spike).
The g-function is defined by the same Thouless-type formula with respect to the
equilibrium measure (cf. section 2(iii)). 
It satisfies the same conditions as in [KMM] 
(measure reality and variational inequality) on bands
(where the branch of the field turns out to be irrelevant) and
on gaps (where the inequalities  are
satisfied according to the branch of the external field).

We have sketched  a  proof  that 
the solution of the initial value problem for the
focusing NLS equation  admits a global (in $x,t$) finite genus representation, 
asymptotically as $h \to 0$, at least in the case of the simplest data 
$u(x,0)=A sechx.$  But the arguments above also hold for
a large class of "semiclassical soliton ensembles" initial data 
defined precisely in section VIII of [KR].

It is worth noting here the recent paper
[LM] which also adresses the issue of the target 
contour hitting the eigenvalue barrier. 
([LM] does not really  mention a variational
problem and prefers to consider directly the conditions
(equations and inequalities) for the g-function.)
This very interesting paper does
not prove the existence of an appropriate
target contour but instead contains a numerical 
and theoretical discussion of the
issue of the target contour hitting the eigenvalue barrier. 

In [LM] the "band" part of the contour is defined not as the support of
an equilibrium measure but instead it is considered as the trajectory
of a quadratic differential (as in [KMM]). It is noted numerically
that such a trajectory may hit the barrier $[0,iA]$. It is 
then proposed that
the  inequalities defining the "gap" part of the target contour   be  amended
when   it passes the barrier $[0,iA]$
and the actual amendment is justified numerically and theoretically.
As a conclusion it is claimed that the mechanism of the second 
"caustic" (a caustic appears when
the topology of $\hat F$ changes as $x,t$ vary)  is different 
from the mechanism of the first caustic.

We  simply note here that the extra
conditions suggested in [LM] appear naturally in the framework we have
introduced above.  The different gap inequalities in [LM]
correspond to gap inequalities in different sheets, as viewed from our 
perspective.  We thus prefer to say that 
the mechanism for  any "caustic" (caused by the change of
the topology of $\hat F$ as $x,t$ vary)  is essentially the same,
independently of whether the maximizing contour
has crossed the spike $[0,iA]$ or not. In any case
the real issue here is not whether we have a first or
second or higher order caustic, but whether
the maximizing contour has  hit the spike $[0,iA]$.
It so happens that for the very specific data $Asechx$ 
the second caustic appears after the maximizing contour
has crossed the spike $[0,iA]$ once, 
but in general this does not have to be the 
case.\footnote{Also, it is  worth recalling  Chapter 6
of [KMM] where it is noticed that, depending on the choices
of the parameters $\sigma $ and $J$ defined there,
the maximizing contour may hit the barrier $[0,iA]$ even
before the first caustic, even for the
simplest $Asechx$ data. So the time where the barrier is first hit is definitely
not an intrinsic parameter of the original problem.}

\bigskip
\newpage

6. CONCLUSION

\bigskip

In the asymptotic analysis of Riemann-Hilbert problems arising 
from integrable systems where the associated Lax operator
is non-self-adjoint, the computation of non-trivial steepest descent contours
is essential. The two main components of a rigorous proof
of asymptotic formulae are:

(i) Proof of the existence and regularity of such steepest descent contours.

(ii) Given (i), a rigorous proof of the asymptotic validity of the deformation 
of the given  Riemann-Hilbert problem
to one with jumps across the  steepest descent contour.

In this review paper we have presented some methods and
results, contained in [KR] and [KMM], achieving
(i) and (ii) for some specific cases of
the initial value problem for the focusing integrable nonlinear
Schr\"odinger equation in the semiclassical limit. We expect that 
these  methods and results may be useful in the
treatment of Riemann-Hilbert problems arising in the analysis
of general complex or normal random matrices [WZ].
Although there have been simple cases of 
non-self-adjoint problems where the target contour 
can be explictly computed without any recourse to a variational problem 
(which of course is always
there; see e.g. [K96], [KSVW], [TVZ]), we believe that global results  can in general
only be justified by proving
existence and regularity for a solution of a 
maximin variational problem in two dimensions.

\bigskip

7. REFERENCES 

\bigskip

[BC] R.Beals, R.Coifman,
Scattering and inverse scattering for first order systems.
Comm. Pure Appl. Math. v.37 (1984), no. 1, pp.39--90. 

[CG] K.F.Clancey, I.Gohberg,
Factorization of matrix functions and singular integral operators.
Operator Theory: Advances and Applications, 3. 
Birkh\"auser Verlag, Basel-Boston, Mass., 1981. 
 
[DVZ94] P.Deift, S.Venakides, X.Zhou,
The Collisionless Shock Region for the Long-Time
Behavior of Solutions of the KdV equation, Comm. Pure Appl. Math., 
v.47, 1994, pp.199-206.

[DVZ97] P.Deift, S.Venakides, X.Zhou, 
New Results in Small Dispersion  KdV  by an Extension of the Steepest 
Descent Method for Riemann-Hilbert Problems, Inter.Math.Res.Notices, 1997, pp.286-299.

[DZ]  P.Deift, X.Zhou, A Steepest Descent Method for
Oscillatory Riemann-Hilbert Problems, Annals of Mathematics, v.137, n.2, 1993,
pp.295-368.

[DZ95] P.Deift,  X.Zhou,
Asymptotics for the Painlev\'e II equation,
Comm. Pure Appl. Math. v.48, 1995 , no. 3, pp.277--337.
 
[E] A.Erd\'elyi, Asymptotic Expansions, Dover 1956.

[I] A.R.Its,
Asymptotics of Solutions of the Nonlinear Schr\"odinger equation, 
and Isomonodromic Deformations of Systems of Linear Differential Equations, Sov.
Math.Dokl., v.24, n.3, 1982, pp.14--18.

[IN] A.R.Its, V.Novokshenov, The Isomonodromic Deformation Method 
in the Theory of Painlev\'e Equations,
Lecture Notes in Math., 1191, Springer Verlag, Berlin 1986.

[K07] S.Kamvissis, in preparation.

[K96] S.Kamvissis, Long time behavior for the focusing nonlinear Schroedinger equation with real 
spectral singularities,  Comm. Math. Phys.  v.180, no. 2, 1996, pp.325--341.

[KMM] S.Kamvissis, K.McLaughlin, P.Miller, Semiclassical Soliton Ensembles
for the Focusing Nonlinear Schr\"odinger Equation, 
Annals of Mathematics Studies, v.154, Princeton University Press, 2003.  

[KR] S.Kamvissis, E.Rakhmanov, Existence and Regularity for an 
Energy Maximization Problem in Two Dimensions,  
Jour. Math. Phys., v.46, n.8, 083505, 2005.

[K00] A.Kuijlaars,  On the finite-gap ansatz in the continuum limit of the Toda lattice,
Duke Math. J.  v.104, no. 3, 2000, pp.433--462.

[KSVW] A.Kuijlaars, H.Stahl, W.Van Assche, F.Wielonsky,  
Asymptotique des approximants de Hermite-Pad\'e
quadratiques de la fonction exponentielle et 
probl\`e\-mes  de Riemann-Hilbert, CRAS v.336, 2003, pp.893-896.

[LL] P.D.Lax, C.D.Levermore, The Zero Dispersion Limit for the KdV Equation, I-III,
Comm. Pure Appl. Math., v.36, 1983, pp.253-290, pp.571-593, 
pp.809-829.

[LM] G.Lyng, P.Miller, The N-soliton of the focusing nonlinear Schr\"odinger 
equation for N large, Comm. Pure Appl. Math., v.60, 2007, pp. 951-1026.

[TVZ] A.Tovbis, S.Venakides, X.Zhou, 
On semiclassical (zero dispersion limit) solutions of the focusing nonlinear 
Schr\"odinger equation, Comm. Pure Appl. Math. 57, n. 7, 2004, pp.877--985. 

[WZ] P.Wiegmann, A.Zabrodin, Large scale correlations in normal non-Her\-mitian matrix ensembles,
J. Phys. A  v.36,  no. 12, 2003, pp.3411--3424.

[Z] X.Zhou,  The Riemann--Hilbert problem and inverse scattering,
SIAM J. Math. Anal., v.20, n.4, 1989, pp. 966--986. 

\enddocument